\documentclass[12pt, draftclsnofoot, onecolumn]{IEEEtran}
\makeatletter
\def\endthebibliography{%
	\def\@noitemerr{\@latex@warning{Empty `thebibliography' environment}}%
	\endlist
}
\makeatother

\usepackage{ifpdf}

\usepackage{cite}

\usepackage{graphicx}
\usepackage{epstopdf} 
\usepackage{xcolor}
\ifCLASSINFOpdf

\fi

\usepackage[cmex10]{amsmath}
\usepackage{amsfonts,amssymb}

\usepackage{algorithmic}

\usepackage{array}

\ifCLASSOPTIONcompsoc
\usepackage[caption=false,font=normalsize,labelfont=sf,textfont=sf]{subfig}
\else
\usepackage[caption=false,font=footnotesize]{subfig}
\fi

\usepackage{dblfloatfix}

\usepackage{url}

\usepackage{arydshln}
\usepackage{enumerate}

\usepackage[mathcal]{euscript}
\newtheorem{theorem}{Theorem}
\newtheorem{lemma}{Lemma}

\newtheorem{corollary}[theorem]{Corollary}

\newtheorem{remark}{\textbf{Remark}}
\newcommand{\vect}[1]{\mathbf{#1}}

\newcommand{\bs}[1]{\boldsymbol{#1}}

\def\beq{\begin{equation}}
\def\eeq{\end{equation}}

\graphicspath {{Figures/}}

\begin{document}
	
	\bstctlcite{IEEEexample:BSTcontrol}
	%
	\title{Humans and Machines can be Jointly Spatially Multiplexed by Massive MIMO}
	%
	%
	%
	
	\author{Kamil~Senel,~\IEEEmembership{Member,~IEEE,}
	Emil~Bj\"{o}rnson,~\IEEEmembership{Senior Member,~IEEE,}
	and~Erik~G.~Larsson,~\IEEEmembership{Fellow,~IEEE}
	\thanks{Parts of this work were presented at the IEEE International Conference on Acoustics, Speech and Signal Processing (ICASSP), 2018 \cite{senel2018icassp}.}
	\thanks{The authors are with the Department of Electrical Engineering (ISY), Link\"{o}ping University, Sweden.}
	\thanks{This work was supported by ELLIIT and the Swedish Foundation for Strategic Research (SSF).}}

	\maketitle

	\begin{abstract}
 
 Future cellular networks are expected to support new communication paradigms such as machine-type communication (MTC) services along with conventional human-type communication (HTC) services. This requires base stations to serve a large number of devices in relatively short channel coherence intervals, which renders allocation of orthogonal pilot sequences per-device in each cell impractical. Furthermore, the stringent power constraints, place-and-play type connectivity and various data rate requirements of MTC devices make it impossible for the traditional cellular architecture to accommodate MTC and HTC services together. Massive multiple-input-multiple-output (mMIMO) technology has the potential to allow the coexistence of HTC and MTC services, thanks to its inherent spatial multiplexing properties and low transmission power requirements. 
 In this work, we first tackle the optimal non-orthogonal pilot design problem and demonstrate that the optimal pilot sequences are Welch bound equality sequences. In the second part, we investigate the performance of a single cell under a shared
 physical channel assumption for MTC and HTC services and propose a novel scheme for sharing the time-frequency resources. The analysis reveals that mMIMO can significantly enhance the performance of such a setup and allow the inclusion of MTC services into the cellular networks without requiring additional resources.  
	\end{abstract}
	

	%
	\IEEEpeerreviewmaketitle

	\section{Introduction} \label{sec:Intro}
	
	\IEEEPARstart{O}{ne} of the key technologies of $5$G future networks is the machine-type communications (MTC), which is projected to provide wireless connectivity to tens of billions of new devices as a result of smart cities, factories, vehicles, and even common objects with sensing and communicating capabilities \cite{dawy2017toward}. A potential solution for accommodating the emerging traffic is utilizing the already existing infrastructure of cellular networks which can provide wide area coverage. The standardization 
	of techniques for MTC over cellular networks is already being considered \cite{3rdGPP}. However, the existing cellular network architectures, which are optimized to handle human-type communications (HTC), must be modified in order to handle MTC alongside HTC, which requires consideration of a diverse communication characteristics \cite{deliverable2015d6}. 
	
	There are crucial problems that must be considered to achieve successful integration of MTC services into the existing cellular networks.~In particular, $5$G networks will have to support a large number of devices with low-complexity constraints and provide various data rates ranging from nearly zero up to multiple gigabits per second with reliability for services that have stringent latency constraints such as health-care, security, and automotive applications \cite{osseiran2014scenarios}.
	
	Another important problem is the pilot shortage problem in MTC. Future networks are expected to support unprecedented number of devices which makes it impossible to assign orthogonal pilots to each active device in the cell. The problem differs from the pilot contamination problem in multi-cell setups in the sense that the contamination is due to devices within the same cell. 	
	The performance of massive MIMO systems under pilot shortage has been investigated in \cite{de2017random} where each device transmits a randomly chosen orthogonal pilot sequence. However, the assumption that pilot sequences must be orthogonal is strictly suboptimal, which will be demonstrated in this work, for MTC (especially for the massive MTC setup in \cite{de2017random}). The optimal design of non-orthogonal pilots in massive MIMO is also considered in \cite{wang2015design}, \cite{akbar2016multi}. However, in this work, we consider a generalized case without relying on fixed power assumption \cite{wang2015design} or asymptotic analysis \cite{akbar2016multi}. 
	
	The performance of cellular networks in a setup where HTC and MTC services coexist has been considered in \cite{dawy2017toward,bontu2014wireless}.~WiFi-based networks constitute a competitive option to cellular networks and the integration of MTC services into the existing WiFi-based networks has been investigated in \cite{sutton2017harmonising}. A potential alternative is to utilize  multihop short-range transmission
	technologies \cite{centenaro2016long}. However, initial experimentations reveal the limitation of short-range technologies for MTC applications and emphasized the requirement of a plug-and-play type of connectivity without centralized planning which can be satisfied by long-range technologies \cite{biral2015challenges}.    
	
	A key technology of $5$G future cellular networks is mMIMO in which the BSs are equipped with a large number of antennas, which gives them the ability to spatially multiplex multiple users \cite{redbook}. The mMIMO technology has been shown to enhance the performance of cellular networks in terms of spectral efficiency for broadband HTC setups and device detection in MTC setups \cite{jiang2016novel,de2017random,liu2018sparse}. However, to the best of authors' knowledge, this is the first work which considers the coexistence of HTC and MTC devices in a mMIMO setup and analyze their joint spectral efficiency and show that mMIMO enabled cellular networks can handle MTC along with HTC without requiring additional resources.

	\subsection{Main Contributions}
	
	In this work, we first address the non-orthogonal pilot sequence design problem and demonstrate that the optimal pilot sequences are Welch bound equality (WBE) sequences. Furthermore, we investigate the performance, in terms of spectral efficiency, of a mMIMO network that concurrently serves devices that utilize HTC and MTC. Different schemes for allocating time-frequency resources between MTC and HTC devices are proposed and compared.~A novel resource allocation scheme is proposed and compared with the orthogonal and non-orthogonal resource allocation schemes. 
	 In particular, we answer the following questions:    
	\begin{itemize}
		\item What is the optimal pilot design for MTC with massive number of devices?
		\item How will the existing cellular networks be affected by the dense MTC deployments?
		\item Can the challenges to accommodate MTC services over cellular networks be handled by the mMIMO technology?
		\item Does the mMIMO technology enable the use of non-orthogonal resources for MTC and HTC services thanks to its inherent utilization of spatial multiplexing? 
	\end{itemize}
	
	 This paper goes beyond the conference version given in \cite{senel2018icassp} which does not consider pilot design problem and only considers resource allocation schemes with random pilot assignment. 
	 
	\section{System Setup}\label{sec:SystemSetup}
\begin{figure}[tb]
		\begin{center}
			\includegraphics[trim=0cm 0cm 0cm 0cm,clip=false, scale = .4]{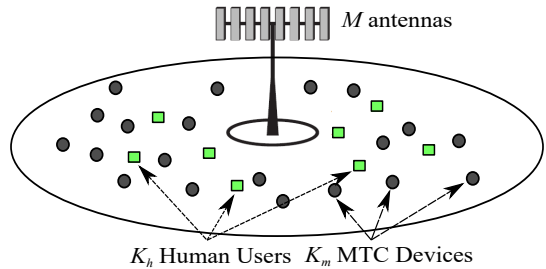}
			\caption{An $M$-antenna base station serves $K$ users, of which $K_m$ are machines and $K_h$ are humans.}
			\label{fig:fig1-SystemSetup} 
		\end{center}
	\end{figure}

	We consider the uplink of a single-cell mMIMO system where a BS with $M$ antennas is serving $K$ single-antenna devices. An example setup is illustrated in Fig.~\ref{fig:fig1-SystemSetup}. There are two types of devices based on the communication they require. Among these devices $K_m$ of them, referred to as \textit{machines}, require machine-type communication and the remaining $K_h = K - K_m$ devices, referred to as \textit{humans}, generate human-type traffic. Humans are assumed to be smaller in numbers and require higher data rates compared to machines. 	 
	
   The time-frequency resources are divided into coherence intervals (CI), such that each channel is constant and frequency-flat in each interval \cite{redbook}. In each CI, the channels take independent realizations from stationary ergodic processes. In the massive MIMO context, a CI consists of the following phases: uplink training, uplink data transmission. The downlink data transmission can either take place in the same CI (as in time-division duplex) or in other dedicated CIs (as in frequency-division duplex).   
   In this paper, we focus on the uplink training and uplink data transmission, while the downlink data transmission analysis is left as future work. 
   Each CI has length $N$ (in samples) and a fraction of these samples are reserved for training whereas the remaining ones are utilized for uplink data transmission. The allocation of samples varies among the schemes considered and the details are provided in Section \ref{sec:Schemes}.

	Non-line-of-sight communication is assumed and the channel between device $k$ and the BS is modeled as
	\begin{equation}
	\vect{g}_k = \sqrt{\beta_k}\vect{h}_k, \forall k = 1,\ldots,K, 
	\end{equation} 
	where $\beta_k$ is the large-scale fading and $\vect{h}_k$ is the small-scale fading. Each element of $\vect{h}_k$ is modeled as i.i.d.$~\mathsf{CN}(0,1)$. The large-scale fading coefficients are assumed to be identical across antennas and known at the BS as they usually change very slowly which makes it possible to acquire accurate estimates. However, the small-scale fading coefficients change independently between CIs and are to be estimated in each CI via uplink training. 

	\subsection{CI Allocation Schemes} \label{sec:Schemes}
	
\begin{figure}[tb]
		\begin{center}
			\includegraphics[trim=0cm 0cm 0cm 0cm,clip=false, scale = .4]{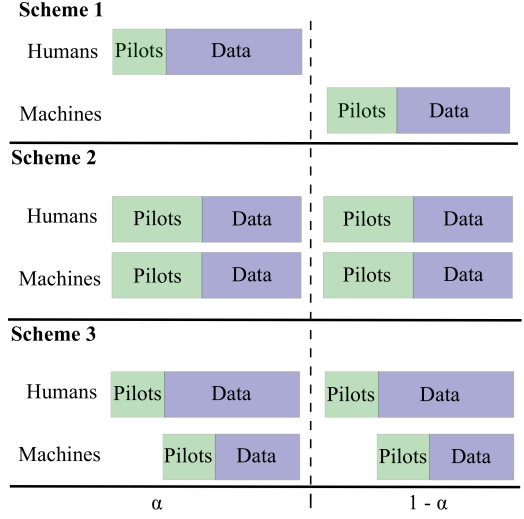}
			\caption{Coherence interval structure for training and data transmission for three different schemes. Here, $\alpha$ and $1-\alpha$ represent the fraction of CIs allocated to humans, and machines in Scheme $1$, respectively. In other schemes, both humans and machines utilize same CIs.}
			\label{fig:CIs} 
		\end{center}
	\end{figure}

	We consider three training and data transmission schemes.
	\begin{itemize}
		\item \textbf{Scheme 1:} Humans and machines utilize different CIs.
		\item \textbf{Scheme 2:} All devices use the same training interval and data transmission interval in every CI. 
		\item \textbf{Scheme 3:} Machines are not allowed to transmit during the training period of humans, which reduces the human's pilot length. After the training of humans, machines transmit their pilot sequences followed by data transmission.   
	\end{itemize}
	Fig.~\ref{fig:CIs} illustrates the CI structures for the three schemes.  Scheme $1$ is an orthogonal allocation scheme in the sense that it allocates different CIs to humans and machines. Scheme $2$ and $3$ are non-orthogonal schemes where both machines and humans utilize the same CIs. In Scheme $3$, we propose a novel approach by utilizing the training period of machines for data transmission of humans, which effectively reduces the pilot overhead for humans. 
	%
	
\section{Channel Estimation and Pilot Design}\label{sec:chEstimationPilotDesign}

In conventional mMIMO setups, all $K$ devices concurrently transmit their pilot sequences and BS estimates the channels based on the received signal. The estimates acquired via this process, referred as uplink training, are utilized to design combining vectors for uplink data transmission (and precoding vectors for downlink data transmission).  

The pilot sequences are generally assumed to be mutually orthogonal for users within a cell. These assumptions on pilot sequences are not realistic for some scenarios in future wireless networks, such as massive MTC with a large number of devices, since it requires many pilots and cumbersome access procedures for pilot assignment \cite{de2017random}. Moreover, constraints on the uplink power budget and excessive overhead signaling compel the use of non-orthogonal pilots for MTC \cite{dawy2017toward,de2017randomM}.   

	In this work, it is assumed that the humans require higher data rates and are smaller in numbers compared to machines, i.e., $K_h < K_m$. As a result, the humans have the privilege to orthogonal pilots, whereas it is not possible to assign orthogonal pilot to machines due to the large number of machines. Moreover, machines are assumed to be low-powered, and low complexity devices which require lower rates. Hence, allocating orthogonal resources to machines requires excessive overhead signaling\footnote{There are exceptions to the low data rate assumption of machines, such as surveillance applications \cite{ratasuk2012coverage}. In these cases the machines with high data rate requirements may be treated as humans.}.
	 
	 Let $\sqrt{N_p}\bs{\varphi}_k \in \mathbb{C}^{N_p \times 1}$ denote the $N_p$-length pilot signal for the $k$th device with $\|\bs{\varphi}_k\|^2 = 1$. It is assumed that humans are always allocated orthogonal pilots. Hence,  
\begin{equation}\label{eq:PilotsHuman}
\bs{\varphi}_k^H\bs{\varphi}_i = 0,~~~ \forall k \in \{1, \ldots, K_h\}, \forall i \in \{1, \ldots, K\}\setminus\{k\}, 
\end{equation}
with SC-$2$ and 
\begin{equation}\label{eq:PilotsHuman2}
\bs{\varphi}_k^H\bs{\varphi}_i = 0,~~~ \forall k \in \{1, \ldots, K_h\}, \forall i \in \{1, \ldots, K_h\}\setminus\{k\}, 
\end{equation}
with SC-$1$ and SC-$3$. 

For machines, non-orthogonal pilots are utilized and we investigate the problem of designing optimal non-orthogonal sequences in terms of channel estimation error. In SC-$1$ and SC-$2$, the active devices concurrently transmit their pilot sequences and the composite received signal at the BS is 
\begin{equation}\label{eq:recSig-sc1}
\vect{Y} = \sqrt{N_p}\sum\limits_{k' \in \mathcal{K}_1} \sqrt{q_{k'}}\vect{g}_{k'} \bs{\varphi}_{k'}^H + \vect{Z}, 
\end{equation}
in SC-$1$. Here, $\vect{Z} \in \mathbb{C}^{M \times N_p}$ is the noise matrix with i.i.d.$~\mathsf{CN}(0,\sigma^2)$ elements and $q_{k'}$ denotes the transmission power of pilot symbols for user $k'$. The set of active devices $\mathcal{K}_1$ is either equal to $\mathcal{K}_m$ or $\mathcal{K}_h$, i.e., the set of machines or humans. In SC-$2$ the received signal at the BS is 
\begin{equation}\label{eq:recSig-sc2}
\vect{Y} = \sqrt{N_p}\sum\limits_{k' \in \mathcal{K}_h} \sqrt{q_{k'}}\vect{g}_{k'} \bs{\varphi}_{k'}^H + \sqrt{N_p}\sum\limits_{k' \in \mathcal{K}_m} \sqrt{q_{k'}}\vect{g}_{k'} \bs{\varphi}_{k'}^H + \vect{Z}, 
\end{equation}
 In SC-$3$, the humans transmit data while machines are training and the composite received signal at the BS during the training of machines is 
\begin{equation}\label{eq:recSig-sc3}
\vect{Y} = \sum\limits_{k' \in \mathcal{K}_h} \hspace{-0.2cm}\vect{g}_{k'}\vect{x}_{k'}^H +    \sqrt{N_p^m}\hspace{-0.2cm}\sum\limits_{k' \in \mathcal{K}_m} \hspace{-0.2cm}\sqrt{q_{k'}}\vect{g}_{k'} \bs{\varphi}_{k'}^H +  \vect{Z}
\end{equation}
where $\vect{x_k} = \sqrt{p_k}\vect{s_k}$ and each element of $\vect{s_k}$ is a unit power symbol to be conveyed by device $k$. In this work, our focus is on the design of optimal non-orthogonal pilot sequences and we leave the joint pilot and data transmit power control problem as future work. Under this assumption, the first term in \eqref{eq:recSig-sc3} can be treated as additive (but not Gaussian) noise. This allows us focus on the training of machines with non-orthogonal pilots and to generalize \eqref{eq:recSig-sc1}, \eqref{eq:recSig-sc2}, and \eqref{eq:recSig-sc3} as 
\begin{equation}\label{eq:recSig-gen}
\vect{Y} = \sqrt{N_p}\sum\limits_{k' \in \mathcal{K}} \sqrt{q_{k'}}\vect{g}_{k'} \bs{\varphi}_{k'}^H + \tilde{\vect{Z}}, 
\end{equation}   
where the set of active devices depends on the scheme. 
    
In order to estimate the channel of device $k \in \mathcal{K}_m$, the BS performs a de-spreading operation on the received signal:
\begin{eqnarray}
\vect{y}_k &=& \vect{Y}\bs{\varphi}_k, \nonumber \\ 
&=& \sqrt{N_p q_k} \vect{g}_k + \sqrt{N_p}\hspace{-0.2cm}\sum\limits_{k' \in \mathcal{K}_m\backslash\{k\}} \hspace{-0.2cm}\sqrt{q_k}\vect{g}_{k'} \bs{\varphi}_{k'}^H\bs{\varphi}_k  + \vect{z}', \nonumber \\
&=&\hspace{-2mm} \sqrt{N_p \beta_kq_k} \vect{h}_k \hspace{-1mm}+ \hspace{-1mm} \sqrt{N_p}\hspace{-0.4cm}\sum\limits_{k' \in \mathcal{K}_m\backslash\{k\}} \hspace{-0.4cm}\sqrt{\beta_{k'}q_{k'}}\vect{h}_{k'} \bs{\varphi}_{k'}^H\bs{\varphi}_k \hspace{-1mm}+ \vect{z}' , \label{eq:despred-1}
\end{eqnarray} 
where $\vect{z}' = \tilde{\vect{Z}}\bs{\varphi}_k$ has i.i.d.$~\mathsf{CN}(0,\sigma^2)$ elements, since $\|\bs{\varphi}_k\|^2 = 1$, for SC-$1$ and SC-$2$. For SC-$3$, $\vect{z}'$ has i.i.d. elements, however it is not necessarily Gaussian. Then, the BS either utilizes
the least squares (LS) estimator or linear minimum mean-square error (LMMSE) estimator to obtain the channel estimate $\hat{\vect{h}}_k$. 
The LS estimate for device $k$ is, 
\begin{eqnarray}
\hat{\vect{h}}^{\textrm{LS}}_k &=& \vect{y}_k(\sqrt{N_p \beta_kq_k})^{-1}, \nonumber \\
&=&  \vect{h}_k + \frac{\sum_{k' \in \mathcal{K}_m\backslash\{k\}} \sqrt{N_p\beta_{k'}q_{k'}}\vect{h}_{k'} \bs{\varphi}_{k'}^H\bs{\varphi}_k  + \vect{z}'}{\sqrt{N_p \beta_kq_k}}.
\end{eqnarray}
The estimation error for device $k$ is 
\begin{eqnarray}
\tilde{\vect{h}}_k^{\textrm{LS}} &=& \hat{\vect{h}}^{\textrm{LS}}_k -  \vect{h}_k, \\
&=& \frac{\sum_{k' \in \mathcal{K}_m\backslash\{k\}} \sqrt{N_p\beta_{k'}q_{k'}}\vect{h}_{k'} \bs{\varphi}_{k'}^H\bs{\varphi}_k  + \vect{z}'}{\sqrt{N_p \beta_kq_k}}.
\end{eqnarray}
The mean square of the $m$th element of $\tilde{\vect{h}}_k^{\textrm{LS}}$ with respect to the small-scale fading coefficients, $\vect{h}$ and noise, $\vect{z}$, is given by
\begin{eqnarray}\label{eq:expectedLSerror}
e_k^{\textrm{LS}} &=& \mathbb{E}_{\vect{h},\vect{z}}\left[\left|\left[\tilde{\vect{h}}_k^{\textrm{LS}}\right]_m\right|^2\right] \\ &=&  \frac{N_p\sum_{k' \in \mathcal{K}_m\backslash\{k\}} \beta_{k'}q_{k'}\left|\bs{\varphi}_{k'}^H\bs{\varphi}_k\right|^2  + \sigma^2}{N_p \beta_kq_k } .	 	
\end{eqnarray}

\begin{remark}
	The expectation in \eqref{eq:expectedLSerror} can also be taken with respect to the pilot sequences which allows us to consider random schemes such as the ones presented in \cite{senel2018icassp,de2017random} where each device chooses one of the $N_p$ orthogonal pilot sequences randomly.
	In these cases, the mean square of the $m$th element of $\tilde{\vect{h}}_k$ with respect to the small-scale fading coefficients, $\vect{h}$, noise, $\vect{z}$ and the set of pilot sequences, $\bs{\phi}$, is given by
\begin{eqnarray}
e_k^{\textrm{LS}} &=& \mathbb{E}_{\bs{\phi}}\left[\mathbb{E}_{\vect{h},\vect{z}}\left[\left|\left[\tilde{\vect{h}}_k^{\textrm{LS}}\right]_m\right|^2|\bs{\phi}\right]\right] \\ &=& \mathbb{E}_{\bs{\phi}}
\left[ \frac{N_p\sum_{k' \in \mathcal{K}_m\backslash\{k\}} \beta_{k'}q_{k'}\left|\bs{\varphi}_{k'}^H\bs{\varphi}_k\right|^2  + \sigma^2}{N_p \beta_kq_k }	\right], \\
& =& 
\frac{N_p\sum_{k' \in \mathcal{K}_m\backslash\{k\}} \beta_{k'}q_{k'}\mathbb{E}_{\bs{\phi}}\left[\left|\bs{\varphi}_{k'}^H\bs{\varphi}_k\right|^2\right]  + \sigma^2}{N_p \beta_kq_k } \label{eq:expectedLSerrorRandom}.	 	
\end{eqnarray}
 These random pilot allocation approaches have desirable advantages for massive MTC setups such as being simple and not requiring any coordination between devices. However, these approaches, as we will demonstrate later, are strictly suboptimal in terms of channel estimation performance.  
 \end{remark}

Next, we consider the LMMSE estimator which is widely used in the massive MIMO context \cite{redbook}. 
The channel estimate given by the LMMSE estimator based on \eqref{eq:despred-1} is as follows:
\begin{equation}\label{eq:MMSEest}
\hat{\vect{h}}_k^{\textrm{LMMSE}} = \frac{\sqrt{N_p \beta_kq_k}}{N_p\sum\limits_{k' \in \mathcal{K}_m}\beta_{k'}q_{k'}|\bs{\varphi}_{k'}^H\bs{\varphi}_{k}|^2+ \sigma^2}\vect{y}_k. 
\end{equation}  
The LMMSE estimate $\hat{\vect{h}}_k^{\textrm{LMMSE}}$ has $M$ i.i.d. elements and the mean-square of the $m$th component is 
\begin{align}\label{eq:MSgamma}
\gamma_{k} = \mathbb{E}_{\vect{h},\vect{z}}\left[\left|\left[\hat{\vect{h}}_k^{\mathrm{LMMSE}}\right]_m\right|^2\right] = \frac{N_p\beta_kq_k}{N_p\sum\limits_{k' \in \mathcal{K}_m}\beta_{k'}q_{k'}|\bs{\varphi}_{k'}^H\bs{\varphi}_{k}|^2+ \sigma^2},
\end{align} 
where $\mathbb{E}_{\vect{h},\vect{z}}$ denotes the expectation with respect to $m$th component of $\vect{h}$ and $\vect{z}$. Let $\tilde{\vect{h}}_k^{\mathrm{LMMSE}} = \hat{\vect{h}}_k^{\mathrm{LMMSE}} - \vect{h}_k$ denote the channel estimation error of device $k$ and the mean-square estimation error of the $m$th component is given by 
\begin{eqnarray}\label{eq:errorMMSE}
e_k^\mathrm{LMMSE} = \mathbb{E}_{\vect{h},\vect{z}}\left[\left|\left[\tilde{\vect{h}}_k^{\mathrm{LMMSE}}\right]_m\right|^2\right] = 1 - \frac{N_p\beta_kq_k}{ N_p\sum\limits_{k' \in \mathcal{K}_m}\beta_{k'}q_{k'}|\bs{\varphi}_{k'}^H\bs{\varphi}_{k}|^2+ \sigma^2}.
\end{eqnarray} 
For the cases where humans transmit data during machines training (SC-$3$), the LMMSE estimator is no longer the true MMSE estimator, however, the succeeding analysis is still valid for all the schemes considered.

We consider the problem of min-max estimation error optimization and aim to find the optimal pilot sequences for machines which provide the min-max solution  \begin{equation}\label{eq:optimalError}
e^* = \min_{\vect{q},\bs{\phi}}\max_{k \in \mathcal{K}_m} ~e_k,
\end{equation}  
where the minimization is with respect to the pilot transmission powers, $\vect{q} = [q_1, \ldots, q_{K_m}]^T$, and pilot sequences, $\bs{\phi}$. First, we investigate the case without any power constraints, i.e., each element of $\vect{q}$ is only assumed to be non-negative. The case with power constraints is analyzed in Section \ref{sec:optPower}. The set of pilot sequences considered, $\bs{\phi}$ contains vectors of the form  $\sqrt{N_p}\bs{\varphi}_k$ with, where $\bs{\varphi}_k \in \mathbb{C}^{N_p \times 1}$ and $\|\bs{\varphi}_k\|^2 = 1$ for all $k \in \mathcal{K}_m$.  

	First note that, $e^*$ is achieved when $e_1 = e_2 = \ldots = e_K$ which can be proved as follows. Suppose $e^*$ can only be achieved when devices have different $e_k$'s and consider a case where $e_1 < e_2 = \ldots, = e_K = e^*$ for a given pilot sequence. Then, by reducing the transmit power, $q_1$, of device $1$, it is possible to obtain a smaller or equal error for other devices which results in $e^* = e_1 = e_2 = \ldots = e_K$. This contradicts the initial assumption that $e^*$ can only be achieved when devices have different channel estimation errors.
	
	Consider the least-squares estimator error given by \eqref{eq:expectedLSerror} which can be rewritten in vector notation as       
	\begin{equation}\label{eq:errorLS_InVect}
	\vect{\Phi}\bs{\mu} + \bs{\eta} = e \bs{\mu},
	\end{equation} 
	where 
	\begin{equation}
	\vect{\Phi} = \begin{bmatrix}
	0 & \left|\bs{\varphi}_{2}^H\bs{\varphi}_1\right|^2 & \ldots & \left|\bs{\varphi}_{K}^H\bs{\varphi}_1\right|^2 \\
	\left|\bs{\varphi}_{1}^H\bs{\varphi}_2\right|^2 & 0 & \ldots & \left|\bs{\varphi}_{K}^H\bs{\varphi}_2\right|^2 \\
	\vdots & & \ddots& \vdots\\
	\left|\bs{\varphi}_{1}^H\bs{\varphi}_K\right|^2&\ldots & & 0
	\end{bmatrix},
	\end{equation}
	and 	
	\begin{equation}\label{eq:mu}
	\bs{\mu} = [\mu_1, \dots, \mu_K]^T,
	\end{equation}
	with $\mu_j = \beta_jq_j$ for all $j \in \mathcal{K}_m$. $e$ is the squared error aimed for each device and the normalized noise vector is  \begin{equation}\label{eq:eta}
	\bs{\eta} = \frac{\sigma^2}{N_p} \vect{1},
	\end{equation}
	where $\vect{1}$ is the all ones vector. Re-writing \eqref{eq:errorLS_InVect} as
	\begin{equation}
	\frac{1}{1 + e}\bar{\vect{\Phi}}\bs{\mu} + \bar{\bs{\eta}}  = \bs{\mu},
	\end{equation}
	where $\bar{\vect{\Phi}} = \vect{\Phi} + \vect{I}$ and $\bar{\bs{\eta}} = \bs{\eta}/(1+e)$, the minimum power solution for a given $e$ is 
	\begin{equation} \label{eq:minPowLS}
	\bs{\mu}^* = \left(\vect{I} - \frac{1}{1 + e}\bar{\vect{\Phi}}\right)^{-1}\bar{\bs{\eta}},
	\end{equation} 
	and there exist a positive $\bs{\mu}$ such that the mean square channel estimation error is $e$ for each device if and only if the spectral radius of $\bar{\vect{\Phi}}$, denoted by $\rho(\bar{\vect{\Phi}})$, is less than $1 + e$ \cite{pillai2005perron}. Note that, it is assumed that there are no constraint on pilot transmit powers to make the problem tractable (Similar result may be obtained by assuming the thermal noise is negligible) and the investigation under a setup with power constraints is presented in Section \ref{sec:optPower}. Based on \eqref{eq:minPowLS}, the following can be stated.
	\begin{lemma}\label{lem:minError}
		The min-max total squared channel estimation error, $e^*$ is obtained when $\rho(\bar{\vect{\Phi}})$ is minimized with respect to $\bs{\varphi}_1, \ldots,\bs{\varphi}_K$. 
	\end{lemma}
	
	An important observation is that $e^*$ is independent of the pilot transmission powers and only depends on the expected correlation between the pilot sequences. Next, we investigate the minimum spectral radius of $\bar{\vect{\Phi}}$. 
	
	$\bar{\vect{\Phi}}$ is a non-negative matrix by definition and hence $\rho(\bar{\vect{\Phi}})$ is an eigenvalue of $\bar{\vect{\Phi}}$ \cite[{Theorem 8.3.1}]{horn1990matrix}. Furthermore, it is a symmetric matrix and a bound on its spectral radius is given by the following lemma.
	\begin{lemma}[Theorem 3.2 in \cite{bo2000spectral}] \label{lem:spectralRadiusBound}
		Let $\vect{A}$ be an $L \times L$ non-negative symmetric matrix. Then 
		\begin{equation}\label{eq:specRadiusBound-1}
		\rho(\vect{A}) \geq \sqrt{\frac{\sum_{i = 1}^{L}d_i^2}{L}},
		\end{equation}
		where $d_j$ is the $j$th row sum of $\vect{A}$ and the equality is achieved when $\vect{A}$ has equal row and column sums.  
	\end{lemma} 
	
	Using the Jensen's inequality, Lemma \ref{lem:spectralRadiusBound} can be extended as follows
	\begin{equation}\label{eq:specRadiusBound-2}
	\sqrt{\frac{\sum_{i = 1}^{L}d_i^2}{L}} \geq \frac{1}{L}\sum_{i = 1}^{L}d_i
	\end{equation}
	with equality if and only if $d_1 = d_2 \ldots, = d_n$. Using the bounds given in \eqref{eq:specRadiusBound-1} and \eqref{eq:specRadiusBound-2} on $\bar{\vect{\Phi}}$, we obtain
	\begin{equation}\label{eq:longInequalityWelch}
	\rho(\bar{\vect{\Phi}}) \geq \sqrt{\frac{\sum_{i = 1}^{K_m}\left(\sum_{j = 1}^{K_m}|\bs{\varphi}_i^H\bs{\varphi}_j|^2\right)^2}{K_m}}\geq \frac{1}{K_m}\sum_{i = 1}^{K_m}\sum_{j = 1}^{K_m}|\bs{\varphi}_i^H\bs{\varphi}_j|^2 \geq \frac{K_m}{N_p},
	\end{equation}
	where the last inequality follows from the Welch bound \cite{welch1974lower}, defined in Appendix \ref{sec:WBEapp}. Any set of vectors satisfying Welch bound is known as Welch bound equality sequences (WBE). Furthermore, any WBE sequence, $\bs{\varphi}_1, \ldots, \bs{\varphi}_{K_m}$ in $\mathbb{C}^{N_p}$, has the following property \cite{waldron2003generalized}
	\begin{equation}
	\sum_{j = 1}^{K_m}|\bs{\varphi}_i^H\bs{\varphi}_j|^2 = \frac{K_m}{N_p}, ~~\forall i = 1, \ldots, K_m.
	\end{equation} 
	Hence, WBE sequences satisfy the inequalities in \eqref{eq:longInequalityWelch} with equality and provides the minimum $\rho(\bar{\vect{\Phi}}) = K_m/N_p$. Based on Lemma \ref{lem:minError}, it can be concluded that WBE sequences minimizes the min-max total squared channel estimation error for the LS estimator. Furthermore, a similar analysis based on \eqref{eq:expectedLSerrorRandom} reveals that WBE sequences also minimize the min-max total squared channel estimation error for random pilot allocation schemes with the LS estimator.

	 
%

	Similarly, the mean square error for the LMMSE estimator given by \eqref{eq:errorMMSE} can be rewritten in vector notation as follows
	\begin{equation}\label{eq:errorMMSE_vect}
	\left(1 - e\right)\left(\bar{\vect{\Phi}}\bs{\mu} + \bs{\eta}\right)  = \bs{\mu},
	\end{equation} 
	and the minimum power solution is 
	\begin{equation}\label{eq:minPowMMSE}
	\bs{\mu}^* = \left(\vect{I} - \left(1 - e\right)\bar{\vect{\Phi}}\right)^{-1}\left(1 - e\right)\bs{\eta}.
	\end{equation}
	Similar to the LS case, there exist a positive $\bs{\mu}$ such that the mean square channel estimation error is $e$ for each device if and only if $\rho(\bar{\vect{\Phi}})$, is less than $1/\left(1 - e\right)$ \cite{pillai2005perron}. The rest of the analysis follows the same steps given in LS case and is therefore omitted. A crucial difference with the MMSE estimator case is that only deterministic pilot allocation schemes are considered, i.e., $\mathbb{E}\left[|\bs{\varphi}_i^H\bs{\varphi}_j|^2\right] = |\bs{\varphi}_i^H\bs{\varphi}_j|^2$. Hence, the performance of random allocation schemes with MMSE estimator are investigated numerically.  	

    The analysis provided above allows us to state the following result regarding the optimal pilot sequences. 
\begin{theorem}\label{thm:th1}
	Consider a system with a set of $N_p$-length pilot sequences $\varphi_1, \ldots, \varphi_{K_m}$ for training where $K_m \geq N_p$ and the channel estimates are acquired via LS or LMMSE estimators. Then, the set of pilot sequences that minimizes the maximum mean-square error if the pilot sequences satisfy the Welch bound with equality.  
\end{theorem}

     Theorem \ref{thm:th1} considers pilot sequences with length $N_p \leq K_m$ and when $K_m = N_p$ the pilot sequences becomes orthogonal. The definition of Welch bound and sequences satisfying the bound is given in Appendix \ref{sec:WBEapp}.   

	The design of WBE codes is fairly straightforward and extensively investigated in the literature \cite{xia2005achieving}. An example of a structured WBE codebook is given in \cite{hochwald2000systematic}, which utilizes normalized $N_p$ distinct rows of a $K_m \times K_m$ FFT matrix. The pilot sequence for device $k$ is
    \begin{equation}\label{eq:WBEsequence}
    \bs{\varphi}_k = \frac{1}{\sqrt{N_p}}\begin{bmatrix}
    e^{j\frac{2\pi}{K_m}u_1 \left(k -1\right)}\\
    e^{j\frac{2\pi}{K_m}u_2 \left(k -1\right)}\\
    \vdots\\
    e^{j\frac{2\pi}{K_m}u_{N_p} \left(k -1\right)}
    \end{bmatrix}
    \end{equation} 
    where $\vect{u} = \left[u_1,\ldots, u_{N_p}\right]^T$ is a vector consisting of $N_p$ parameters is to be selected. A simple choice of $u_i = i$ for all $i = 1,\ldots,N_p$ provides a set of WBE sequences without any consideration of min-max correlation. Hence, assuming that the set of active devices is known, each device can generate its pilot sequence using \eqref{eq:WBEsequence}.

\subsection{MMSE vs LS}

Next we compare the performance of the two estimation techniques under a setup where WBE sequences are utilized as pilots. Based on the analysis above, we can state the following. 
\begin{lemma}\label{lem:limitPerformanceWBE}
	Consider a system with $K_m$ users and $N_p$-length WBE pilot sequences used for uplink training. Then, the achievable  min-max channel estimation error, $e^{\mathrm{LMMSE}}$, with the LMMSE estimator is bounded as
	\begin{equation} \label{eq:minMMMSEerr}
	e^{\mathrm{LMMSE}} \geq \frac{K_m - N_p}{K_m},
	\end{equation}
	and the error, $e^{\mathrm{LS}}$, with the LS estimator is bounded as  
	\begin{equation}\label{eq:minLSerr}
	e^{\mathrm{LS}} \geq \frac{K_m - N_p}{N_p},
	\end{equation} 
	for $N_p \leq K_m$.  
\end{lemma}
\begin{IEEEproof}
	Recall that it is possible to achieve the mean square channel estimation error, $e$, for each device if and only if $\rho(\bar{\vect{\Phi}})$, is less than $1/\left(1 - e\right)$ with the LMMSE estimator. Then, we have
	\begin{eqnarray}
	\rho(\bar{\vect{\Phi}}) &\leq& \frac{1}{1 - e^{\mathrm{LMMSE}} }, \\ 
	1 - \frac{1}{\rho(\bar{\vect{\Phi}})} &\leq&       e^{\mathrm{LMMSE}}. 
	\end{eqnarray} 
	The lower bound for $e^{\mathrm{LMMSE}}$ is achieved when  $\rho(\bar{\vect{\Phi}})$ is minimized and the minimum $\rho(\bar{\vect{\Phi}}) = K_m/N_p$, which is given in \eqref{eq:longInequalityWelch}. Hence, the minimum $e^{\mathrm{LMMSE}}$ given by \eqref{eq:minMMMSEerr} is achieved when $\rho(\bar{\vect{\Phi}}) = K_m/N_p$. Similarly, for the LS estimator case, we have
	\begin{eqnarray}
	\rho(\bar{\vect{\Phi}}) &\leq& 1 + e^{\mathrm{LS}}, \\ 
	\rho(\bar{\vect{\Phi}})-1 &\leq&       e^{\mathrm{LS}}. 
	\end{eqnarray} 
	Using, the minimum $\rho(\bar{\vect{\Phi}}) = K_m/N_p$, \eqref{eq:minLSerr} can be obtained which concludes the proof. 
\end{IEEEproof}

\begin{figure}[tb] 
	\begin{center}
		\includegraphics[trim=0cm 0cm 0cm 0cm,clip=false, scale = .6]{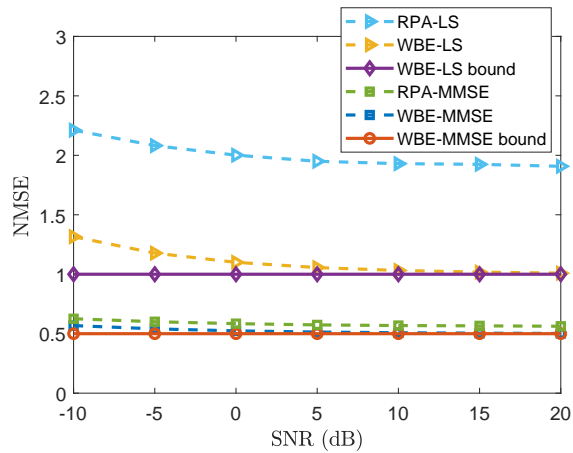}
		\caption{Normalized mean channel estimation error as a function of SNR, for pilot sequence length, $N_p = 10$, number of machine devices, $K_m=20$ and number of BS antennas, $M = 50$. }
		\label{fig:fig1} 
	\end{center}
\end{figure}

The bounds given by Lemma \ref{lem:limitPerformanceWBE} provides a performance limits for WBE sequences. The bounds can only be achieved as $\mathrm{SNR} \rightarrow \infty$. In Fig.~\ref{fig:fig1}, the normalized mean squared error (NMSE) of channel estimates, $\mathbb{E}\{\|\vect{h}_k - \hat{\vect{h}}_k \|^2\}/M$, for WBE sequences and RPA scheme is depicted with LMMSE and LS estimators. In this particular example, the pilot sequence length is $N_p = 10$ and $K_m = 20$. As expected, the LMMSE estimator provides better performance than the LS estimator for both pilot sequences. An interesting result is that the RPA scheme performs close to the optimal WBE sequences with LMMSE, especially at higher SNRs which suggests it might be a good allocation scheme for massive MTC where ability to operate without coordination is a desirable property. However, results presented based on channel estimation errors for a particular case should not be used to draw general conclusions. 

\begin{figure}[tb] 
	\begin{center}
		\includegraphics[trim=0cm 0cm 0cm 0cm,clip=false, scale = .6]{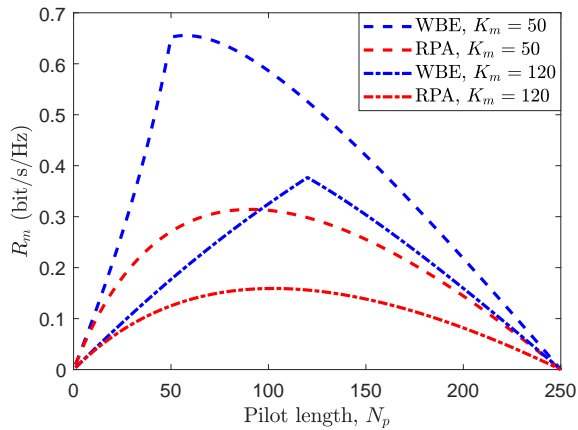}
		\caption{Achievable rate as a function of machine pilot length for different number of devices, under a setup with number of BS antennas, $M = 500$ and coherence interval length, $N = 250$. }
		\label{fig:fig2} 
	\end{center}
\end{figure}

Fig.~\ref{fig:fig2} provides a comparison of RPA and WBE based pilot sequences in terms of achievable ergodic rates. In this example, two cases with different number of machines are considered and the curves depict the achievable rates based on maximum ratio combining. The rate expression and its derivation are detailed in Section \ref{sec:SC-1Analysis}. The results suggests that it is possible to provide a given data rate to a higher number of machines by assigning them WBE pilot sequences instead of utilizing RPA scheme.  
The performance difference between different pilot sequences is most significant when $N_p \approx K_m$ which is expected as at $N_p = K_m$ WBE sequences become orthogonal sequences.

\subsection{Optimal Transmit Powers}\label{sec:optPower}

The analysis provided so far demonstrates that the optimal pilot sequences are WBE sequences, however WBE sequences are not unique for a given $N_p$ and $K_m$. Next, we consider the transmit powers required for WBE sequences and state the following. 
\begin{lemma}\label{lem:samePow}
	Consider a system with a set of $N_p$-length WBE sequence utilized for training where $K_m \geq N_p$ and let $q^*$ denote the minimum power required for a given $e$. Then, any WBE sequence can provide the same $e$ with identical $q^*$.
\end{lemma} 
\begin{IEEEproof}
	Consider two different the power vectors $\vect{q}_1^*$ and $\vect{q}_2^*$ for two WBE sequences which provides the same $e$. The minimum power vectors are given by \eqref{eq:minPowLS}:
	\begin{equation}
	\vect{q}_i^* = \vect{B}^{-1}\left(\vect{I} - \frac{1}{1 + e}\bar{\vect{\Phi}}_i\right)^{-1}\bar{\bs{\eta}},~~\text{for}~~i = 1,2.
	\end{equation}	
	Even though, $\bar{\vect{\Phi}}_1$ and $\bar{\vect{\Phi}}_2$ are not necessarily identical, their row and column sums are identical. Hence, the inverse $\left(\vect{I} - \frac{1}{1 + e}\bar{\vect{\Phi}}_1\right)^{-1}$ and $\left(\vect{I} - \frac{1}{1 + e}\bar{\vect{\Phi}}_2\right)^{-1}$ also have identical row sums and
	\begin{equation}
	\left(\vect{I} - \frac{1}{1 + e}\bar{\vect{\Phi}}_1\right)^{-1}\bar{\bs{\eta}} = \left(\vect{I} - \frac{1}{1 + e}\bar{\vect{\Phi}}_2\right)^{-1}\bar{\bs{\eta}}
	\end{equation}  
	which concludes the proof.
\end{IEEEproof}

Lemma \ref{lem:samePow} demonstrates that the required power vectors are identical for any WBE sequence. The result can further be extended to provide a closed form expression for the minimum power vector, $\vect{q}^* = [q_1^*, \ldots, q_{K_m}^*]^T$ as follows:
\begin{equation} \label{eq:optPowLS}
q_k^* = \frac{\sigma^2}{\left(N_p\left(1 + e\right) - K_m\right)\beta_k}, ~~~ \text{for}~k = 1,\ldots,K_m, 
\end{equation}	
with the LS estimator and
\begin{equation}\label{eq:optPowMMSE}
q_k^* = \frac{\sigma^2\left(1-e\right)}{\left(N_p - K_m\left(1 - e\right)\right)\beta_k}, ~~~ \text{for}~k = 1,\ldots,K_m, 
\end{equation}	
with the LMMSE estimator. Equations \eqref{eq:optPowLS} and \eqref{eq:optPowMMSE} shows that the transmission powers depend on the pilot length, number of users, target channel estimation error and large-scale fading coefficients. Since, $N_p$, $K_m$ and $e$ are identical for each user, the difference between the transmission powers are due to $\beta_k$'s. In a practical system with power constraints, if a user is not able to provide the required power, then either, $N_p$ must be increased or the system should be adjusted for a higher $e$.

	\subsection{Pilot Transmission Power Control}
	
	In this part, we investigate the pilot transmission power control problem for both humans and machines\footnote{ 
	We leave the joint pilot and data transmit power control problem as future work and assume that power control during uplink training and data transmission are disjoint problems.}. It is assumed that humans transmit with maximum power during uplink training, which is reasonable since they are allocated orthogonal pilots and there is no incentive to utilize any power level below the maximum. However, this is not the case for the machines.     
	
	In order to determine a realistic power control strategy for MTC, different  constraints (based on the MTC scenario) must be considered. For example, in ultra-reliable MTC (uMTC) which requires reliable communication with low latency, elaborate power control schemes may be employed whereas simple power control strategies are suitable for MTC scenarios with low-complexity, low-power devices. Especially, for the mMTC uplink, complex power control approaches based on small-scale	fading coefficients are not practical as accurate channel state information can only be acquired by allocating additional resources. Furthermore, mMTC devices usually require low bit-rates and transmit small packages which makes elaborate power control techniques redundant for these low-complexity devices.   
	
	The results of \eqref{eq:optPowLS} and \eqref{eq:optPowMMSE} suggests that the transmit power of machines should scale inversely proportional to their large-scale fading coefficients. Therefore, we employ statistical channel inversion (SCI) power control for machines during training. SCI is a power control technique that only relies on the large-scale fading and helps to combat the near-far effect \cite{massivemimobook}. Such an approach especially benefits devices with weaker channel gains. In SCI, the devices adjust their powers as follows: 
	\begin{flalign} \label{eq:SCI} 
	q_k = q_{ul}^{\max} \frac{\beta_{\min}}{\beta_k},
	\end{flalign}
	where $\beta_{\min}$ represents the large-scale fading coefficient of a device at cell edge and $q_{ul}^{\max}$ denotes the maximum transmission power. With SCI, each device's transmission power scale inversely proportionally with respect to their large-scale coefficients during uplink training.

	\section{Achievable Rate Analysis}\label{sec:SpectralAnalysis}
	
	In this section, the achievable rates of the three schemes illustrated in Fig.~\ref{fig:CIs} are investigated. Each scheme has an uplink training phase followed by data transmission. Although we consider the case where the channel estimates are acquired via LMMSE estimator, the extension to LS estimators or other alternative estimators is straight-forward. 
	
	\subsection{Analysis of Scheme 1} \label{sec:SC-1Analysis}
	
	In Scheme $1$ (SC-$1$), humans and machines utilize different CIs which prevents any interference between them. The active devices 
	concurrently transmit their pilot sequences and the channel estimates of active devices are acquired. In SC-$1$ either humans or machines are active i.e., $\mathcal{K} = \mathcal{K}_h$ or $\mathcal{K} = \mathcal{K}_m$.
		
	Let $s_k$ denote the unit power symbol to be conveyed by device $k$. Then, device $k$ transmits $x_k = \sqrt{p_k} s_k$,
	where $p_k$ is the data transmit power of device $k$. In order to detect the data symbols of the $k$th device, the BS employs the maximum ratio combining (MRC) with the combining vector
	\begin{equation}\label{eq:precoder}
	\hat{\vect{v}}_k = \frac{1}{\gamma_k\sqrt{M}}\hat{\vect{h}}_k
	\end{equation}
	to compute the inner product with the received signal, 
	\begin{equation}
	\vect{y} = \sum_{k' \in \mathcal{K}}\vect{g}_{k'}x_{k'} + \vect{z}
	\end{equation} as 
	\begin{eqnarray}\label{eq:compositeSignalatBS}
	y_k = {\vect{v}}_k^H\vect{y} 
	&=&   \sum_{k' \in \mathcal{K}} {\vect{v}}_k^H\vect{g}_{k'}x_{k'} + {\vect{v}}_k^H\vect{z}.
	\end{eqnarray} 
	Based on \eqref{eq:compositeSignalatBS}, the achievable rate of device $k$ is given by
	\begin{equation}\label{eq:SEofHumanSC-0}
	R_k = \alpha_k\left( \frac{N_d}{N} \right)\log_2\left(1+ \Gamma_k\right)	
	\end{equation}
	where the effective SINR term in \eqref{eq:SEofHumanSC-0} is given by \cite{medard2000effect}  
	\begin{equation} \label{eq:useForgetBound}
	\Gamma_k = \frac{|\mathbb{E}\{y_ks_k^*\}|^2}{\mathbb{E} \{|y_k|^2\} - |\mathbb{E}\{y_ks_k^*\}|^2 }, 
	\end{equation}
	and
	\begin{equation}\label{eq:CIdivisionSC-0}
	N_d = \begin{cases}
	N- N_p^h,  &\text{if}~~ k \in \mathcal{K}_h,\\
	N - N_p^m, &\text{if}~~ k \in \mathcal{K}_m.
	\end{cases}
	\end{equation}
	Here, $N_p^h$ and $N_p^m$ are the pilot lengths for humans and machines, respectively. $\alpha_k \in [0,1]$ represents the fraction of CIs assigned to humans/machines. For SC-$1$, we have the following result.
	
		\begin{lemma}\label{lem:SEofSC-1}
		The achievable rate of device $k$ under Scheme $1$ is 
		\begin{equation}\label{eq:SEofHumanSC-1}
		R_k =  \alpha_k\left(\frac{N_d}{N} \right)\log_2\left(1+ \Gamma_k\right)	
		\end{equation}
		where $\Gamma_k$ is the effective SINR for device $k$ and is given by  
	\begin{equation}\label{eq:SEofMachineSC-0}
\Gamma_k = \begin{cases}
\frac{M\beta_{k}p_k}{ \frac{1}{\gamma_{k}}\left( \sum\limits_{k' \in \mathcal{K}_h}p_{k'}\beta_{k'} + \sigma^2 \right) }, &\text{if}~~ k \in \mathcal{K}_h,\\
\frac{M\beta_{k}p_k}{\frac{1}{\bar{\gamma}_{k}}\left( \sum\limits_{k' \in \mathcal{K}_m}p_{k'}\beta_{k'} + \sigma^2 \right) + M\hspace{-0.3cm} \sum\limits_{k' \in \mathcal{K}_m^k}\hspace{-0.2cm}\frac{p_{k'}q_{k'}\beta_{k'}^2\mathbb{E}_{\bs{\phi}}\left[\left|\bs{\varphi}_{k'}^H\bs{\varphi}_k\right|^2\right] }{q_k\beta_{k}} }, &\text{if}~~ k \in \mathcal{K}_m,
\end{cases}
\end{equation}
where $\mathcal{K}_m^k = \mathcal{K}_m \backslash\{ k \}$ and
\begin{equation} 
\bar{\gamma}_k = \mathbb{E}_{\bs{\phi}}\left\lbrace\frac{1}{\gamma_{k}}\right\rbrace^{-1}\hspace{-.25cm} = \frac{N_p^mq_k\beta_k}{ N_p^m\sum\limits_{k' \in \mathcal{K}_m}q_{k'}\beta_{k'}\mathbb{E}_{\bs{\phi}}\left[\left|\bs{\varphi}_{k'}^H\bs{\varphi}_k\right|^2\right]  + \sigma^2}, \forall k \in \mathcal{K}_m. \nonumber 
\end{equation} 
	\end{lemma}
\begin{IEEEproof}
	See Appendix \ref{sec:lem1Proof}.
\end{IEEEproof}

Note that in \eqref{eq:SEofMachineSC-0}, the effective SINR with random pilot allocations is also considered as the expectation can be taken with respect to pilot sequences. There is no interference between humans and machines since they are served orthogonally. Furthermore, since humans are assigned orthogonal pilots, there is no pilot contamination and therefore, no coherent interference, i.e., interference that scales with the number of antennas, between humans. However, there is coherent interference between machines as a result of non-orthogonal pilots.

	
	\subsection{Analysis of Scheme 2} \label{sec:Scheme-1SE} 
	In SC-$2$, each device uses $N_p$ symbols for training and $N-N_p$ symbols for data. To find the corresponding rate of device $k$, we utilize the bounding techniques given in \cite{redbook} and state the following:
	\begin{lemma}\label{lem:SEofSC-2}
		The achievable rate of device $k$ under Scheme $2$ is 
		\begin{equation}\label{eq:SEofHumanSC-2}
		R_k = \left( \frac{N - N_p}{N} \right)\log_2\left(1+ \Gamma_k\right)	
		\end{equation}
		where $\Gamma_k$ is the effective SINR for device $k$ and is given by  
		\begin{equation}\label{eq:SEofMachineSC-1}
		\Gamma_k = \begin{cases}
		\frac{M\beta_kp_k}{ \frac{1}{\gamma_{k}}\left( \sum\limits_{k' \in \mathcal{K}}p_{k'}\beta_{k'} + \sigma^2 \right) }\,, &\text{if}~~ k \in \mathcal{K}_h,\\
		\frac{M\beta_kp_k}{\frac{1}{\bar{\gamma}_{k}}\left( \sum\limits_{k' \in \mathcal{K}}p_{k'}\beta_{k'} + \sigma^2 \right) + M\hspace{-.2cm} \sum\limits_{k' \in \mathcal{K}^k_m}\hspace{-0.25cm}\frac{p_{k'}q_{k'}\beta_{k'}^2\mathbb{E}_{\bs{\phi}}\left[\left|\bs{\varphi}_{k'}^H\bs{\varphi}_k\right|^2\right]}{q_k\beta_{k}} } \,, &\text{if}~~ k \in \mathcal{K}_m,
		\end{cases}
		\end{equation} 
		where 
\begin{equation} 
\bar{\gamma}_k = \mathbb{E}_{\bs{\phi}}\left\lbrace\frac{1}{\gamma_{k}}\right\rbrace^{-1}\hspace{-.25cm} = \frac{N_pq_k\beta_k}{ N_p\hspace{-.2cm}\sum\limits_{k' \in \mathcal{K}_m}\hspace{-.2cm}q_{k'}\beta_{k'}\mathbb{E}_{\bs{\phi}}\left[\left|\bs{\varphi}_{k'}^H\bs{\varphi}_k\right|^2\right]  + \sigma^2}, \forall k \in \mathcal{K}_m. \nonumber 
\end{equation} 
	\end{lemma}
		\begin{IEEEproof} See Appendix \ref{sec:lem2Proof}.
			\end{IEEEproof}
	The effective SINRs given by \eqref{eq:SEofMachineSC-1} reveals that as long as orthogonal pilots are assigned to humans, the integration of machines into an existing network does not create coherent interference to the humans. Hence, as $M$ grows, the effect of the additional interference originating from machines vanishes. However, this is not the case for machines as they suffer coherent interference due to the use of non-orthogonal pilots. Notice that the intra-class coherent interference also depends on the choice of pilot sequences.  
	
	\subsection{Analysis of Scheme 3} \label{sec:Scheme-2SE} 
	In SC-$3$, the machines are silent during the training of humans and send their pilot sequences while humans are transmitting data. This scheme favors humans in the sense that, they start transmitting data immediately after training without considering the training of machines. 
	The LMMSE channel estimate for machines is  
	\begin{equation} \label{eq:ChannelEstimateMachine-SC-2}
	\hat{\vect{h}}_k = \frac{\sqrt{N_p^mq_k\beta_k}}{ N_p^m\hspace{-0.2cm}\sum\limits_{k' \in \mathcal{K}_m}\hspace{-0.2cm}q_{k'}\beta_{k'}|\bs{\varphi}_{k'}^H\bs{\varphi}_{k}|^2 + \hspace{-0.2cm} \sum\limits_{k' \in \mathcal{K}_h}\hspace{-0.2cm}p_{k'}\beta_{k'} + \sigma^2} \vect{y}_k, \quad \forall k \in \mathcal{K}_m,
	\end{equation}
	which is not the MMSE estimator since 
	\begin{equation}
	\vect{y}_k = \sum\limits_{k' \in \mathcal{K}_h} \hspace{-0.2cm}\vect{g}_{k'}\vect{x}_{k'}^H\bs{\varphi}_k +    \sqrt{N_p^m}\hspace{-0.2cm}\sum\limits_{k' \in \mathcal{K}_m} \hspace{-0.2cm}\sqrt{q_{k'}}\vect{g}_{k'} \bs{\varphi}_{k'}^H\bs{\varphi}_k  + \vect{z}'
	\end{equation}
	is not Gaussian due the first term as $\vect{x}_k = \sqrt{p_k}\vect{s}_k$, where each element of $\vect{s}_k$ is a unit power symbol to be conveyed by device $k$. 
	Notice that the human's data symbols transmitted during machine training phase, introduce coherent interference from humans to machines and deteriorate their channel estimation quality.  
	\begin{lemma}\label{lem:SEofSC-3}
		The achievable rate of device $k \in \mathcal{K}_m$ under Scheme $3$ is 
		\begin{equation}\label{eq:SEofMachSC-3}
		R_k = \left(\frac{N - N_p^h-N_p^m}{N} \right)\log_2\left(1+ \Gamma_k\right)	
		\end{equation}
		where 
		 $\Gamma_k$ is given by 
		\begin{equation}\label{eq:SEofMachineSC-3}
		\Gamma_k = 
		\frac{Mp_k\beta_k}{\frac{1}{\bar{\gamma}_k}\left(\sum\limits_{k' \in \mathcal{K}}\hspace{-0.2cm}p_{k'}\beta_{k'} + \sigma^2 \right) + M\left(\hspace{-0.02cm}\sum\limits_{k' \in \mathcal{K}_m^k}\hspace{-0.25cm}\frac{p_{k'}q_{k'}\beta_{k'}^2\mathbb{E}_{\bs{\phi}}\left[\left|\bs{\varphi}_{k'}^H\bs{\varphi}_k\right|^2\right]}{q_k\beta_k} + \hspace{-0.25cm}\sum\limits_{k' \in \mathcal{K}_h}\hspace{-0.2cm}\frac{p_{k'}^2\beta_{k'}^2}{N_p^mq_k\beta_k}\right) },
		\end{equation} 
		and 
		\begin{equation} 
		\bar{\gamma}_k = \mathbb{E}_{\bs{\phi}}\left\lbrace\frac{1}{\gamma_{k}}\right\rbrace^{-1}\hspace{-.25cm} = \frac{N_p^mq_k\beta_k}{ N_p^m\hspace{-.2cm}\sum\limits_{k' \in \mathcal{K}_m}\hspace{-.2cm}q_{k'}\beta_{k'}\mathbb{E}_{\bs{\phi}}\left[\left|\bs{\varphi}_{k'}^H\bs{\varphi}_k\right|^2\right]  + \hspace{-.2cm}\sum\limits_{k' \in \mathcal{K}_h}\hspace{-.2cm}p_{k'}\beta_{k'} + \sigma^2}, \forall k \in \mathcal{K}_m. 
		\end{equation} 
		For humans, $k \in \mathcal{K}_h$, the achievable rate under SC-$3$ is given by\footnote{In the conference version \cite{senel2018icassp}, the provided rate expression of humans for Scheme 3 is only valid under the assumption that machines use the same transmit power during both training and data transmission. Here, we provide a rate expression for humans without any assumptions on the power levels of the machines.}
		\begin{equation}\label{eq:SEofHumanSC-3}
		R_k = \left(\frac{N_p^m}{N} \right)\log_2\left(1+ \Gamma_{k,1}\right) +  \left(\frac{N- N_p^h - N_p^m}{N} \right)\log_2\left(1+ \Gamma_{k,2}\right) 
		\end{equation} 
		where
		\begin{equation}
		\Gamma_{k,1} =\frac{M\beta_kp_k}{\frac{1}{\gamma_{k}}\left( \sum\limits_{k' \in \mathcal{K}_h}\hspace{-0.2cm}p_{k'}\beta_{k'} + \sum\limits_{k' \in \mathcal{K}_m}\hspace{-0.2cm}q_{k'}\beta_{k'} + \sigma^2\right) },~~ \,k \in \mathcal{K}_h,	
		\end{equation}
		and
		\begin{equation}
		\Gamma_{k,2} = \frac{M\beta_kp_k}{\frac{1}{\gamma_{k}}\left( \sum\limits_{k' \in \mathcal{K}}\hspace{-0.2cm}p_{k'}\beta_{k'} + \sigma^2\right)},~~ \,k \in \mathcal{K}_h,	
		\end{equation}
		
	\end{lemma}
\begin{IEEEproof}
	See Appendix \ref{sec:lem3Proof}.
	\end{IEEEproof}
%
	In SC-$3$, humans start data transmission after training without waiting for the training of machines to finish. Hence, the number of available data symbols for humans is higher in SC-$3$ which comes at a cost of causing coherent interference to the machines. This also results in two different terms in \eqref{eq:SEofHumanSC-3}, which corresponds to different achievable rates by humans during the training of machines and data transmission of machines.  
	
	\subsection{Zero-Forcing Receiver}
	
	So far, we have only considered MRC, however, mMIMO provides another linear combining technique, zero-forcing (ZF), which aims to cancel interference between devices. In this part, we assume that ZF is utilized for humans while machines still employ MRC at the receiver and derive ergodic rate expression for the three schemes introduced in Section \ref{sec:Schemes}.
	
	\begin{remark}
	 It should be noted that the rate expressions of the machines are not affected by using ZF receiver for humans, hence the expressions given in previous sections are valid for machines.  
	\end{remark}
	
	During the training phase, humans utilize orthogonal pilots, and after de-spreading of the received composite signal, we have 
	\begin{equation}\label{eq:ZFdespread}
\vect{y}_k =  \sqrt{N_p \beta_kq_k} \vect{h}_k + \vect{z}', \quad \forall k \in \mathcal{K}_h.	
	\end{equation}
	Based on \eqref{eq:ZFdespread}, the LMMSE estimate of device $k$ is 
\begin{equation}\label{eq:ZF-MMSE}
\hat{\vect{h}}_k^{\textrm{LMMSE}} = \frac{\sqrt{N_p \beta_kq_k}}{N_p\beta_{k}q_{k}+ \sigma^2}\vect{y}_k, \quad \forall k \in \mathcal{K}_h,	
\end{equation}
which is identical for all the schemes. The mean-square of the channel estimate is 
\begin{equation}
\gamma_{k} = \mathbb{E}_{\vect{h},\vect{z}}\left[\left|\left[\hat{\vect{h}}_k^{\mathrm{LMMSE}}\right]_m\right|^2\right] = \frac{N_p\beta_kq_k}{N_p\beta_{k}q_{k}+ \sigma^2}, \quad \forall k \in \mathcal{K}_h.
\end{equation}
 Next, we investigate the achievable rates of humans under different schemes. 
 
	\begin{description}
		\item [\textbf{SC-$1$}]: The achievable rate of device $k$ under SC-$1$ is 
		\begin{equation}\label{eq:SEofHumanSC-1zf}
		R_k =  \alpha_k\left(\frac{N_d}{N} \right)\log_2\left(1+ \Gamma_k\right),	
		\end{equation}
		where $\Gamma_k$ is the effective SINR for device $k$ and is given by  
		\begin{equation}\label{eq:SEofHumans-1zf}
		\Gamma_k =
		\frac{\left(M-K_h\right)\gamma_{k}\beta_{k}p_k}{  \sum\limits_{k' \in \mathcal{K}_h}p_{k'}\beta_{k'}\left(1 - \gamma_{k'}\right) + \sigma^2  },\quad   \forall k \in \mathcal{K}_h.
		\end{equation}
		\item [\textbf{SC-$2$}]: The achievable rate of device $k$ under SC-$2$ is 
		\begin{equation}\label{eq:SEofHumanSC-2zf}
		R_k =  \left(\frac{N-N_p}{N} \right)\log_2\left(1+ \Gamma_k\right),	
		\end{equation}
		where $\Gamma_k$ is given by  
		\begin{equation}\label{eq:SEofHumans-2zf}
		\Gamma_k =
		\frac{\left(M-K_h\right)\gamma_{k}\beta_{k}p_k}{  \sum\limits_{k' \in \mathcal{K}_h}p_{k'}\beta_{k'}\left(1 - \gamma_{k'}\right) + \sum\limits_{k' \in \mathcal{K}_m}p_{k'}\beta_{k'} + \sigma^2  },\quad   \forall k \in \mathcal{K}_h.
		\end{equation}
		\item [\textbf{SC-$3$}]: In this scheme, the rate of the humans is given by 
				\begin{equation}\label{eq:SEofHumanSC-3zf}
		R_k = \left(\frac{N_p^m}{N} \right)\log_2\left(1+ \Gamma_{k,1}\right) +  \left(\frac{N- N_p^h - N_p^m}{N} \right)\log_2\left(1+ \Gamma_{k,2}\right), 
		\end{equation} 
		where 
		\begin{equation}
\Gamma_{k,1} =	\frac{\left(M-K_h\right)\gamma_{k}\beta_{k}p_k}{  \sum\limits_{k' \in \mathcal{K}_h}p_{k'}\beta_{k'}\left(1 - \gamma_{k'}\right) + \sum\limits_{k' \in \mathcal{K}_m}q_{k'}\beta_{k'} + \sigma^2  },\quad   \forall k \in \mathcal{K}_h,	
\end{equation}
and
\begin{equation}
\Gamma_{k,2} = 	\frac{\left(M-K_h\right)\gamma_{k}\beta_{k}p_k}{  \sum\limits_{k' \in \mathcal{K}_h}p_{k'}\beta_{k'}\left(1 - \gamma_{k'}\right) + \sum\limits_{k' \in \mathcal{K}_m}p_{k'}\beta_{k'} + \sigma^2  },\quad   \forall k \in \mathcal{K}_h.	
\end{equation}
	\end{description}
The rate expressions can be derived by replacing the terms, $M$ in the coherent gain by $M - K_h$, and replacing $\beta_{k}$ by $\beta_{k}\left(1 - \gamma_{k}\right)$ in the interference term for all $k \in \mathcal{K}_h$. This is analogous to how the expressions for ZF and MRC are related in \cite[Section 3.2]{redbook}.

	\subsection{Asymptotic Analysis}
	\label{sec:Asymptotic}
	In order to gain further insights into the performance of the resource allocation schemes under a massive MIMO setup, the asymptotic limits of the rate expressions as $M \rightarrow \infty$ are investigated in this section. The analysis reveals the limitations of the system due to the coherent interference.
	Note that as $M \rightarrow \infty$, the rate of humans, $R_k \rightarrow \infty$, $\forall k \in \mathcal{K}_h$ in all of the schemes considered. This is to be expected as humans suffer no coherent interference thanks to the orthogonal pilots allocated for them. However, this is not the case for machines and the asymptotic limits are summarized as follows.	
	\begin{corollary}\label{cor:Asymptotic}
		The achievable SINR for device $k \in \mathcal{K}_m$ as $M~\rightarrow \infty$ is given by 
		\begin{equation}\label{eq:SEofMachineAsymptotic}
		\Gamma_k = \begin{cases}
		~~ \frac{\beta_kp_k}{\hspace{-.1cm} \sum\limits_{k' \in \mathcal{K}_m^k}\hspace{-0.05cm}\frac{p_{k'}q_{k'}\beta_{k'}^2\mathbb{E}_{\bs{\phi}}\left[\left|\bs{\varphi}_{k'}^H\bs{\varphi}_k\right|^2\right] }{q_k\beta_{k}}}, &\text{for SC-1 and SC-2}, \\
		\frac{\beta_kp_k}{ \left(\sum\limits_{k' \in \mathcal{K}_m^k}\hspace{-0.25cm}\frac{p_{k'}q_{k'}\beta_{k'}^2\mathbb{E}_{\bs{\phi}}\left[\left|\bs{\varphi}_{k'}^H\bs{\varphi}_k\right|^2\right]}{q_k\beta_k} + \hspace{-0.15cm}\sum\limits_{k' \in \mathcal{K}_h}\hspace{-0.2cm}\frac{p_{k'}^2\beta_{k'}^2}{N_p^mq_k\beta_k}\right) } , &\text{for SC-3.}
		\end{cases}
		\end{equation}
	\end{corollary}  
	
	The proof follows from taking the limit in \eqref{eq:SEofMachineSC-0}, \eqref{eq:SEofMachineSC-1} and \eqref{eq:SEofMachineSC-3}. 
	The asymptotic analysis shows that SC-$1$ and SC-$2$ are equivalent in terms of asymptotic SINR whereas in SC-$3$ machines suffer from additional coherent interference originating from humans. In both cases, the effective SINR increases with the pilot length of machines, which is not necessarily identical for each scheme.    
	

	\section{Numerical Results} 
	
	In this section, numerical results are presented for the schemes introduced in Section \ref{sec:SystemSetup} and analyzed in Section \ref{sec:SpectralAnalysis}. The simulation setup consists of a single cell where the humans and machines are uniformly and independently distributed. The simulation parameters are summarized in Table~\ref{tbl:SysParameters}.
	
	\begin{table}
		\centering
		\caption{Simulation Parameters}
		\label{tbl:SysParameters}
		\begin{tabular}{l|l}
			\hline
			\textbf{System Parameter} & \textbf{Value} \\ \hline
			Path loss at distance $d$ (km) & 130 + 37.6 $\log_{10}(d)$ \\ 
			Cell Radius     & $250\,$m           \\ 
			Minimum Distance ($d_{min}$)          & $20\,$m           \\ 
			Total Noise Power ($B_w \sigma^2$)         & 2$\cdot10^{-13}\,$W \\ 
			Maximum UL-Transmit Power ($\rho_{ul}^{\max}$)         & $1\,$W         \\  
		Number of Humans ($K_h$)        & 5                \\ 
			Number of Machines ($K_m$)        & 45                \\ \hline
			
		\end{tabular}
	\end{table}
	
	%
	
    \begin{figure}[tb]
		\begin{center}
			\includegraphics[trim=.5cm .05cm 0cm 0.6cm,clip=true, width = 9cm,height = 6.2cm]{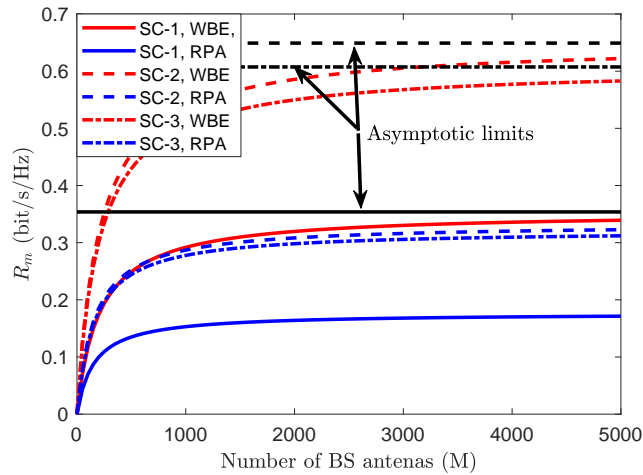}
			\caption{Achievable ergodic rate and asymptotic limits for machines with each scheme respect to number of antennas.}
			\label{fig:AsymptoticAnalysis}\vspace{-6mm}
		\end{center}
	\end{figure}

	The asymptotic limits provided in Corollary \ref{cor:Asymptotic} and the ergodic achievable rate with respect to the number of BS antennas are depicted in Fig.~\ref{fig:AsymptoticAnalysis}. Note that in each of the schemes $R_h \rightarrow \infty$ as $M \rightarrow \infty$ and therefore only the ergodic achievable rates of machines are included in the simulations. The curves are obtained by employing SCI during both training and data transmission. The pilot lengths are optimized at each $M$ for all schemes considered by employing a grid search. For each scheme the WBE sequences give a better performance compared to RPA. Among the schemes considered SC-$2$ provides the best rate as in this scheme all of the coherence intervals are utilized by the machines and there is no coherent interference originating from humans.

	\begin{figure}[tb]
		\begin{center}
			\includegraphics[trim=0cm 0cm 0cm 0cm,clip=true, width = 9cm,height = 6.1cm]{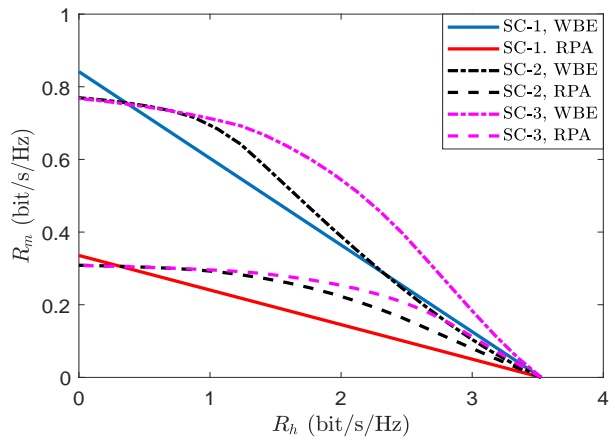}
				\caption{Rate regions for max-min rates obtained via different schemes for $50$ devices, $K_h = 5,~K_m = 45$ with $M = 100$ and coherence interval length, $N = 100$.}
			\label{fig:RateRegions} 	\vspace{-6mm}
		\end{center}
	\end{figure}

	Fig.~\ref{fig:RateRegions} depicts the rate regions for the schemes described in Section~\ref{sec:SpectralAnalysis} for different pilot allocation methods. Here, $R_h$ and $R_m$ denotes the max-min rate for humans and machines respectively. The rate curves are obtained by maximizing the minimum rate with respect to the transmit powers and the machine pilot length $N_p^m$. This optimization problem is solved by formulating it as a geometric programming problem and using CVX to obtain the solution \cite{cvx}. The pilot length of humans is fixed at $N_p^h = K_h$, which creates the difference between schemes when $R_h = 0$. When there is only one type of active device in a given CI, i.e., either only humans or machines, SC-$1$ performs the best. However, for the cases where machines and humans coexist, allowing transmission from both results in a higher ergodic achievable rate as illustrated by the non-orthogonal SC-$2$ and SC-$3$.
	Each scheme shows significant improvements when the training is accomplished via WBE sequences compared to RPA. Among the non-orthogonal schemes, the new proposed SC-$3$ performs the best.

	\begin{figure}[tb]
		\begin{center}
			\includegraphics[trim=0cm 0cm 0cm 0cm,clip=true, width = 9cm,height = 6.3cm]{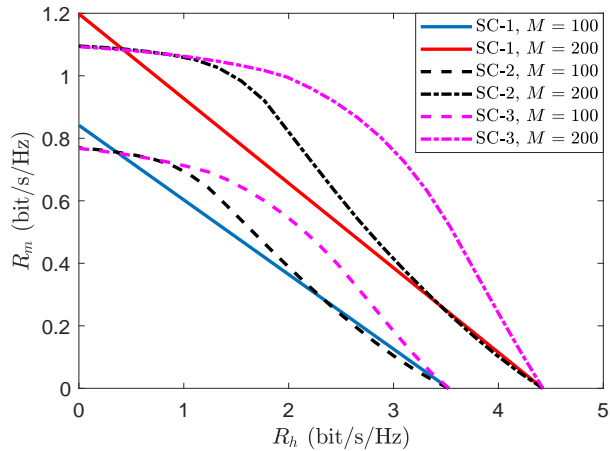} 
			\caption{Rate regions for max-min rates obtained for different number of BS antennas with $K = 50$ devices ($K_h = 5$, $K_m = 45$) and coherence interval length $N = 100$.}
			\label{fig:rateWrtM} \vspace{-6mm}
		\end{center}
	\end{figure} 
	
	The impact of the number of antennas is illustrated in Fig.~\ref{fig:rateWrtM}. For this particular example, the training is carried out using WBE sequences at each of the cases considered. As $M$ increases the non-orthogonal schemes (SC-$2$, SC-$3$) outperform orthogonal scheme (SC-$1$) due to two important reasons. First, the effect of the non-coherent interference between humans and machines decreases with $M$, effectively converging to the SINR in the orthogonal scheme as $M \rightarrow \infty$. Also, the pre-log factor becomes dominant with increasing $M$ as the system starts to operate in the bandwidth limited region at higher $M$ values contrary to lower $M$ values in which the system is in the power limited region. The difference between SC-$2$ and SC-$3$ is due to the fact that once the machines are active in a CI, the humans have to wait for their training with SC-$2$ which degrades the performance of SC-$2$, especially at higher $R_h$.   
	
		\begin{figure}[tb]
		\begin{center}
			\includegraphics[trim=0cm 0cm 0cm 0cm,clip=true, width = 9cm,height = 6.3cm]{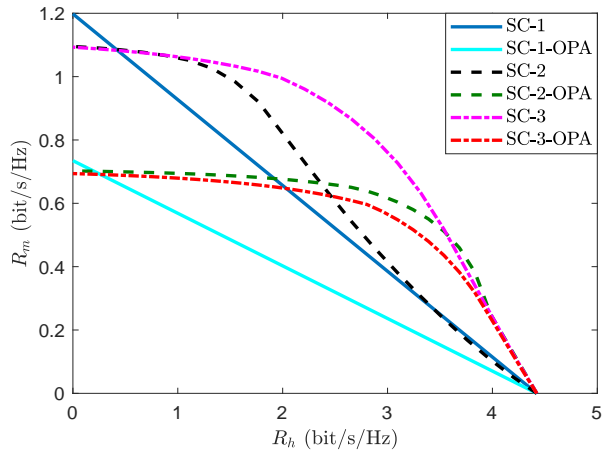} 
		\caption{Comparison of rate regions for max-min rates between orthogonal and non-orthogonal pilot allocation for machines with $K = 50$ devices ($K_h = 5$, $K_m = 45$), $M=200$ BS antennas, and coherence interval length $N = 100$.} \vspace{-6mm}
			\label{fig:rateOrt} 
		\end{center}
	\end{figure} 

    So far, we have assumed that machines are allocated non-orthogonal pilots. However, by scheduling active machines over multiple CIs, it is possible to allocate orthogonal pilots to each device in the system. Fig.~\ref{fig:rateOrt} illustrates the rate regions obtained under an orthogonal pilot allocation (OPA) setup and provides comparison with the non-orthogonal pilot allocation schemes. The orthogonal pilots to machines are allocated as follows, at each CI only a group of machines, consisting of nine devices for this particular example, are active. Hence, over five CIs each device can be served and can utilize orthogonal pilots. We assume that humans are active in all of CIs and they are always allocated orthogonal pilots. The resulting rate regions are depicted in Fig.~\ref{fig:rateOrt}, which are obtained under a setup with $M=200$ BS antennas, and a coherence interval length of $N = 100$. The results reveals that non-orthogonal pilot allocation schemes, especially SC-$3$ performs better than OPA, except for a small region. However, it should be noted that in the simulations the overhead signaling cost required by the OPA due to scheduling is ignored.     

\begin{figure}[tb]
	\begin{center}
		\includegraphics[trim=0cm 0cm 0cm 0cm,clip=true, width = 9cm,height = 6.3cm]{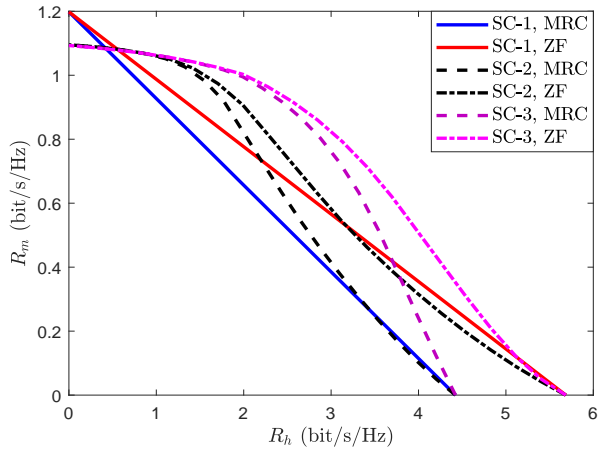} 
		\caption{Comparison of rate regions for max-min rates between MRC and ZF receiver for humans, under a setup with $K = 50$ devices ($K_h = 5$, $K_m = 45$), $M=200$ BS antennas, and coherence interval length $N = 100$.}
		\label{fig:rateZF} \vspace{-6mm}
	\end{center}
\end{figure} 

    Fig.~\ref{fig:rateZF} illustrates the rate regions obtained for the three schemes when different receivers are utilized for humans. The ZF receiver has a better performance compared to MRC, under all of the considered schemes.
    This is to be expected, as using the ZF receiver, humans experience less interference from other humans and hence, can achieve the same spectral efficiency with smaller transmit powers compared to MRC. This in turn, decreases the total interference on machines and results in a higher capacity region for both humans and machines. Also note that among the schemes considered, SC-$3$ provides the best performance.

	%

	\section{Conclusion}
	
	In this work, we consider the problem of accommodating both MTC and HTC in a mMIMO system and proposed a novel resource allocation scheme along with the analysis of achievable rates for HTC and MTC. The characteristic of MTC makes the traditional orthogonal pilot assignment method not feasible and non-orthogonal pilots must be considered. The optimal design of non-orthogonal pilots is addressed in the first part of this work which reveal WBE sequences are optimal in terms of mean-square channel estimation error. The numerical analysis not only validates the theoretical results on the optimality of WBE sequences but also demonstrates that significant performance gains are possible in terms of spectral efficiency using WBE sequences for channel estimation. Moreover, two non-orthogonal schemes (SC-$2$ and SC-$3$) which allows a BS to serve machines and humans simultaneously are considered and compared with an orthogonal scheme (SC-$1$) which serves machines and humans in different coherence intervals. The ergodic spectral efficiency expressions for humans and machines in each scheme are derived. The analysis shows that mMIMO is a key technology for enabling MTC in cellular networks thanks to the large array gain and spatial multiplexing. 
	
	\appendices
	\section{Welch Bound Equality Sequences}\label{sec:WBEapp}
	Let $\vect{b}_1, \ldots, \vect{b}_n$ be a set of unit norm vectors in $\mathbb{C}^d$ where $n \geq d$. Then, a lower bound on the cross correlation of vectors, known as the Welch bound, is given by 
	\begin{equation}\label{eq:WelchBound}
       \sum_{i = 1}^n\sum_{j = 1}^n |\vect{b}_i^H\vect{b}_j|^2 \geq \frac{n^2}{d}, 
	\end{equation}
	and any sequence of vectors which satisfies the bound with equality is called Welch bound equality sequence. 
	
	\section{Proof of Lemma \ref{lem:SEofSC-1}}\label{sec:lem1Proof}
	First, we consider the rate expression for the humans. The terms in \eqref{eq:useForgetBound} are given by 
	\begin{equation}
	\mathbb{E}\left[y_ks_k^*\right] = \sqrt{M\beta_kp_k},~ \quad \forall k \in \mathcal{K}_h,
	\end{equation}
	and 
    \begin{equation}
    \mathbb{E}\left[{|y_k|^2}\right] = \frac{1}{\gamma_k} \left(\sum_{k' \in \mathcal{K}_h}\beta_{k'}p_{k'} + \sigma^2\right) + M\beta_{k}p_k,~ \quad \forall k \in \mathcal{K}_h. 
    \end{equation}   
    Similarly, for the machines, we have
    \begin{equation}
    \mathbb{E}\left[y_ks_k^*\right] = \sqrt{M\beta_kp_k},~ \quad \forall k \in \mathcal{K}_m,
    \end{equation}
    and
    \begin{flalign}
    \mathbb{E}\left[{|y_k|^2}\right] =& \mathbb{E}\left[\left|\sum_{k' \in \mathcal{K}_m}\vect{v}_k^H\vect{h}_{k'}\sqrt{p_{k'}\beta_{k'}}s_{k'} + \vect{v}_k^H\vect{z}\right|^2\right],\\ 
    =& \sum_{k' \in \mathcal{K}_m}\mathbb{E}\left[\left|\vect{v}_k^H\vect{h}_{k'}\right|^2\right]p_{k'}\beta_{k'} + \sigma^2\mathbb{E}\left[\|\vect{v}_k\|^2\right], \\ 
    =& Mp_k\beta_k + \frac{1}{\bar{\gamma}_{k}}\left(\sum_{k' \in \mathcal{K}_m}p_{k'}\beta_{k'} + \sigma^2\right) + M\hspace{-0.2cm} \sum\limits_{k' \in \mathcal{K}_m^k}\hspace{-0.2cm}\frac{p_{k'}q_{k'}\beta_{k'}^2\mathbb{E}_{\bs{\phi}}\left[\left|\bs{\varphi}_{k'}^H\bs{\varphi}_k\right|^2\right] }{q_k\beta_{k}}, \forall k \in \mathcal{K}_m.
    \end{flalign}
    The computation of expectations can be carried out by using techniques introduced in \cite{bjornson2016massive} and \cite{redbook}.
   \section{Proof of Lemma \ref{lem:SEofSC-2}}\label{sec:lem2Proof}
	SC-$2$ rate expressions are very similar to SC-$1$. The difference is that the training and data transmission of both humans and machines are concurrently carried out. The rate expression for the humans can be derived by computing
\begin{equation}
\mathbb{E}\left[y_ks_k^*\right] = \sqrt{M\beta_kp_k},~ \quad \forall k \in \mathcal{K}_h,
\end{equation}
and 
\begin{equation}
\mathbb{E}\left[{|y_k|^2}\right] = \frac{1}{\gamma_k} \left(\sum_{k' \in \mathcal{K}}\beta_{k'}p_{k'} + \sigma^2\right) + M\beta_{k}p_k,~ \quad \forall k \in \mathcal{K}_h. 
\end{equation}   
The terms in \eqref{eq:useForgetBound} for the machines are given as follows,
\begin{equation}
\mathbb{E}\left[y_ks_k^*\right] = \sqrt{M\beta_kp_k},~ \quad \forall k \in \mathcal{K}_m,
\end{equation}
and
\begin{flalign}
\mathbb{E}\left[{|y_k|^2}\right] =& \mathbb{E}\left[\left|\sum_{k' \in \mathcal{K}}\vect{v}_k^H\vect{h}_{k'}\sqrt{p_{k'}\beta_{k'}}s_{k'} + \vect{v}_k^H\vect{z}\right|^2\right],\\ 
=& \sum_{k' \in \mathcal{K}}\mathbb{E}\left[\left|\vect{v}_k^H\vect{h}_{k'}\right|^2\right]p_{k'}\beta_{k'} + \sigma^2\mathbb{E}\left[\|\vect{v}_k\|^2\right], \\ 
=& Mp_k\beta_k + \frac{1}{\bar{\gamma}_{k}}\left(\sum_{k' \in \mathcal{K}}p_{k'}\beta_{k'} + \sigma^2\right) + M\hspace{-0.2cm} \sum\limits_{k' \in \mathcal{K}_m^k}\hspace{-0.2cm}\frac{p_{k'}q_{k'}\beta_{k'}^2\mathbb{E}_{\bs{\phi}}\left[\left|\bs{\varphi}_{k'}^H\bs{\varphi}_k\right|^2\right] }{q_k\beta_{k}}, \forall k \in \mathcal{K}_m.
\end{flalign}
   \section{Proof of Lemma \ref{lem:SEofSC-3}}\label{sec:lem3Proof}
   In SC-$3$, we need to compute two rate expressions for humans, which correspond to the achievable rate during machines training and data transmission. In both cases, we have, 
   \begin{equation}
   \mathbb{E}\left[y_ks_k^*\right] = \sqrt{M\beta_kp_k},~ \quad \forall k \in \mathcal{K}_h.
   \end{equation}
   The achievable rate of humans, during the machines training, i.e., while machines are transmitting their $N_p^m$-length pilots, is given by 
   \begin{equation}
   \mathbb{E}\left[{|y_k|^2}\right] = \frac{1}{\gamma_k} \left(\sum_{k' \in \mathcal{K}_h}\beta_{k'}p_{k'} + \sum_{k' \in \mathcal{K}_m}\beta_{k'}q_{k'} +  \sigma^2\right) + M\beta_{k}p_k,~ \quad \forall k \in \mathcal{K}_h, 
   \end{equation}
   and when both machines and humans are transmitting data, is given by
   \begin{equation}
   \mathbb{E}\left[{|y_k|^2}\right] = \frac{1}{\gamma_k} \left(\sum_{k' \in \mathcal{K}}\beta_{k'}p_{k'} +  \sigma^2\right) + M\beta_{k}p_k,~ \quad \forall k \in \mathcal{K}_h. 
   \end{equation} 
   Note that, since the interference level during human data transmission may change based on whether the machines are transmitting pilots or data, the humans needs to employ two different codebooks with different modulation and coding scheme. Another possibility is to employ identical power control during the pilot and data transmission of machines. 
    
   In SC-$3$, the LMMSE estimate of the machines is not the true MMSE estimate since after de-spreading, the resulting signal is not Gaussian. The terms in \eqref{eq:useForgetBound} for the machines can be computed as follows:
   \begin{equation}
   \mathbb{E}\left[y_ks_k^*\right] = \sqrt{M\beta_kp_k},~ \quad \forall k \in \mathcal{K}_m,
   \end{equation}
   and 
   \begin{flalign}
   \mathbb{E}\left[{|y_k|^2}\right] =& \mathbb{E}\left[\left|\sum_{k' \in \mathcal{K}}\vect{v}_k^H\vect{h}_{k'}\sqrt{p_{k'}\beta_{k'}}s_{k'} + \vect{v}_k^H\vect{z}\right|^2\right],\\ 
   =& \sum_{k' \in \mathcal{K}}\mathbb{E}\left[\left|\vect{v}_k^H\vect{h}_{k'}\right|^2\right]p_{k'}\beta_{k'} + \sigma^2\mathbb{E}\left[\|\vect{v}_k\|^2\right], \\ 
   =& Mp_k\beta_k + \frac{1}{\bar{\gamma}_{k}}\left(\sum_{k' \in \mathcal{K}}p_{k'}\beta_{k'} + \sigma^2\right) + M\hspace{-0.2cm} \sum\limits_{k' \in \mathcal{K}_m^k}\hspace{-0.2cm}\frac{p_{k'}q_{k'}\beta_{k'}^2\mathbb{E}_{\bs{\phi}}\left[\left|\bs{\varphi}_{k'}^H\bs{\varphi}_k\right|^2\right] }{q_k\beta_{k}} \\ +& M \sum_{k' \in \mathcal{K}_h} \frac{p_{k'}^2\beta_{k'}^2}{N_p^m q_k \beta_{k}}, \quad \forall k \in \mathcal{K}_m.
   \end{flalign}

	
	
	%
	\bibliographystyle{IEEEtran}

	
	%
	
	%
	%
	%
	
	
	

\end{document}